\documentclass[aps,prb,12pt,showpacs]{revtex4}
\usepackage{graphicx}
\usepackage[english,russian]{babel}
\usepackage{amssymb}\usepackage{amsmath}
\begin{document}
\title{Comments on the quantum correlation and entanglement}
\author{S. V. Gantsevich and V. L. Gurevich}
\affiliation{Ioffe Institute, Russian Academy of Sciences,
194021 Saint Petersburg, Russia}%
\begin{abstract}
\baselineskip=3.1ex
\indent
In recent decades it was established that the quantum measurements
of physical quantities in space-time points divided by space-like
intervals may be correlated. Though such correlation follows from the
formulas of quantum mechanics its physics so far remains unclear
and there is a number of different and rather contradictory
interpretations. They concern particularly the so-called
Einstein-Podolsky-Rosen paradox where the momentary action at a
distance together with non-local entangled states is used for the
interpretation(see, e.g. [\cite{Cav,NPG,WSG,M,R,AFOV,Zr, RBH,Zel}]).

We assume that the quantum theory can be
formulated as local and look for the consequences of this
assumption. Accordingly we try to explain the correlation
phenomena in a local way looking for the origin of correlation.
To exclude a presupposed correlation of participating
quantum particles we consider two independent particle sources and
two detectors that are independent as well. We show that the
origin of the correlation is the feature that the occupation
number of a particle (and other its measurable quantities) is
formed by a pair of complex conjugated wave functions with in
general arbitrary phases. We consider this point as crucial as it
provides interpretation of the observed correlation phenomena that
may otherwise look puzzling.

We briefly discuss a special type of noise that is typical for
the quantum correlation phenomena.
\end{abstract}
 \vspace{0.7truecm}
\maketitle
\baselineskip=3.5ex
\par
We consider two independent sources of quantum particles (bosons
or fermions) and two independent detectors to register them. We
assume that there is a source $U$ generating particles in the
state $u$ and another source $V$ generating particles in the state
$v$. Let the wave functions of these states be the eigenfunctions
of the time-independent Hamiltonian $H$ so their space shape remains
unchanged during the time evolution
$$\psi(r,t)=\exp(-iHt)\psi(r,0)=\exp(-i\omega t)\psi(r,0).$$ We
assume also that there are two spatially separated detectors
measuring physical quantities $A$ and $B$ and the results of
measurements can be recorded and compared. For simplicity we
assume that the measurements are performed at the same time $t$.
\par
To begin with, let us suppose that the particles from source $U$
can reach detector $A$ while the particles from source $V$ can
reach detector $B$. Then the observable quantities $\bar{A}$ and
$\bar{B}$ are given by
\begin{equation}\label{1}
\bar{A}=\langle u|A|u\rangle F_u, \hspace{1cm} \bar{B}=\langle
v|B|v\rangle F_v.
\end{equation}
Here $F_u$ and $F_v$ are the state occupancies. They
describe the intensities of the sources. Since the source particles do
not interact their occupancies are constant
$$F(t)=F(0).$$
\par
Further on we will use the Dirac notation and name the wave
function $|\dots\rangle$ as {\em ket} while the complex conjugated
wave function $\langle\dots|$ will be named as {\em bra}. Let us
emphasize that to present a result of measurement of a physical
quantity one needs two items,  i. e. a {\em bra} and a {\em ket}.
Thus an observed quantum particle is always represented by a pair
{\em bra}\,+\,{\em ket}. For brevity we will call such a pair as
"wavicle". (This term was sometimes used in the popular literature
in order to accentuate the distinction between quantum particles
and usual waves). We will say that the sources emit wavicles and
the detectors register them. It is important for what follows
that the emitted {\em bra} and {\em ket} have arbitrary initial
phases $\varphi$ but the phase values of the {\em bra} and {\em
ket} of the {the same} source should be {\em the same} (with
the opposite signs). Therefore they cancel in the expressions for
such quantities as $\bar{A}$ and $\bar{B}$ in Eq.~(\ref{1}). The
additional phase factors $e^{i\omega t}$ and $e^{-i\omega t}$
acquired by the {\em bra} and {\em ket} during the time evolution
are also cancelled.
\par
Let us assume that the sources emit particles one by one and
they reach the detectors producing
data. According to the quantum mechanics, the measurements
give eigenvalues of operators $A$ and $B$ with certain
probabilities. For instance, the mean value $\bar{A}$ can be
represented as the sum
\begin{equation}\label{2}
\langle u|A|u\rangle
F_u=F_u\sum_jc_j^{*}(u)c_j(u)A_j=F_u\sum_j|c_j(u)|^2A_j\equiv
F_u\sum_jA_jw_j(u).
\end{equation}
We introduce the functions $f_j$ as the orthogonal and normalized
eigenfunctions of the operator $A$ with eigenvalues $A_j$ so that
$$Af_j=A_jf_j,\quad|u\rangle =\sum_jc_j(u)f_j,\quad w_j(u)=|c_j(u)|^2.$$
The aforementioned initial phases of the
{\em bra} and {\em ket} of the state $u$ do not influence the
probabilities $w_j(u)$ in Eq.~(\ref{2}).
\par
As the next step we assume that the particles from the sources
$U$ and $V$ can reach both detectors, $A$ and $B$. We know that
for a measurement their {\em bra} and {\em ket} should "meet"\, in
the detector. If they both come from the same source one has the
situation of Eq.~(\ref{2}). Then we have for the mean data
in each detector the sum of independent contributions from both
sources:
\begin{equation}\label{3}
\overline{A}=\langle u|A|u\rangle F_u+\langle v|A|v\rangle
F_v,\quad\overline{B}=\langle u|B|u\rangle F_u+\langle
v|B|v\rangle F_v.
\end{equation}
The {\em bra}\, and\, {\em ket} of each source evolve
independently according to their equations of motion but neither
the time factors $e^{\pm i\omega t}$ nor the initial phases enter
into Eq.~(\ref{3}).
\par
The correlation $\overline{AB}$ according to the detector data
of Eq.~(\ref{3}) is:
\begin{equation}\label{4}
\overline{AB}=[\langle u|A|u\rangle \langle v|B|v\rangle + \langle
v|A|v\rangle \langle u|B|u\rangle ]F_u F_v.
\end{equation}
In this expressions a flow of weavicles emitted by the sources
looks as a flow of classical point-like particles that hit both
detectors with the probability $F_u F_v$. We have a U-particle and
a V-particle. One of them hits one detector and another one hits
the other. In this case there is no correlation between the
detector data. (Of course, the sources themselves may be
initially correlated and to describe this phenomenon one should have
replaced the product of independent intensities $F_u F_v$ in
Eq.~(\ref{4}) by the average product $\overline{F_u F_v}$. This
type of correlation is of the classical nature and we do not
consider it).
\par
We remind that weavicles emitted by the sources are
represented by pairs {\em bra} + {\em ket}. Thus we actually have
not two but four objects, i.e. two {\em bra} and two {\em ket}
with two values of initial phases $\pm\varphi_u$ and
$\pm\varphi_v$. We understand that a pair of {\em bra} and {\em ket}
is necessary for the performance of a detector device.
\par
A new and interesting situation emerges where the {\em bra} and
{\em ket} come into detectors pairwise {\em from two different
sources.} In this case the wavicles that hit the detectors are not
the wavicles emitted by a single source. To get these new wavicles
one should exchange either two {\em bra} or two {\em ket} of the
pairs of previous case (\ref{4}).  It is crucial that such {new
weavicles entering both detectors should be inevitably correlated
since they have equal phases $\phi$ of the opposite sign.} Such a
phase is the difference between the initial phases and the phases
acquired by the {\em bra} and {\em ket} during the time evolution:
\begin{equation}\label{5}
\phi=\pm i(\omega_u t-\varphi_u-\omega_v t+\varphi_v).
\end{equation}
Making the exchange procedure in Eq.~(\ref{4})
$$|u\rangle \leftrightarrows|v\rangle\quad \mbox{and}\quad \langle
u|\rightleftarrows\langle v|$$ we get the following correlation
contributions
\begin{equation}\label{6}
(\overline{AB})_{\rm cor}=\pm [\langle u|A|v\rangle \langle v|B|u\rangle +
\langle u|B|v\rangle \langle v|A|u\rangle ]F_uF_v.
\end{equation}
Note that phase $\phi$ does not enter into the final expression for
the average $(\overline{AB})_{\rm cor}$. The correlation contribution is
negative for fermions and positive for bosons. The total average
$\overline{AB}$ is the sum of the uncorrelated and correlated
parts:
\begin{equation}\label{7}
\overline{AB}=[\langle u|A|u\rangle \langle v|B|v\rangle \pm
\langle u|A|v\rangle \langle v|B|u\rangle ]F_u
F_v+(u\rightleftarrows v).
\end{equation}
Here the upper (lower) sign is for bosons (fermions). Let us note
that the correlation contribution reveals itself only after
joint averaging of actual measurements of both detectors. On the
one hand, if one averages the data of each detector separately one
gets Eq.~(\ref{4}) and finds no trace of correlation. On the other
hand, the quantity $\overline{AB}$ is not measured directly but is
obtained by the multiplication of recorded detector data that are
measured independently. It follows that these data should
implicitly contain the correlation contributions that vanish after
the separate averaging but {\em emerge as a result of the joint
averaging.}
\par
The new correlated weavicles correspond to the mixed quantum
states. Let us consider the measurements in mixed quantum states
in more detail and interpret them. The pair of {\em bra} and {\em
ket} taken from different sources comes into the detector with an
arbitrary initial phase difference $\phi$. The contributions from
such pairs are always proportional to a phase factor $e^{i\phi}$.
Though this factor becomes zero after averaging over $\phi$ the
vanishing itself occurs as a result of summation over many
measurement events. By analogy with Eq.~(\ref{2}) for an
observable quantity in a mixed quantum state with the initial
phase difference $\phi$ we have:
\begin{equation}\label{8}
{\rm Re}\langle
u|A|v\rangle=\sum_j|c_j^{*}(u)c_j(v)|A_j\cos[(\alpha_j+\phi)].
\end{equation}
Here $\alpha_j$ is an additional phase difference depending on the
coefficients $c_j$. The quantities $\langle u|A|v\rangle$ and
$\langle v|A|u\rangle=\langle u|A|v\rangle^*$ participate with
equal probabilities so that one can consider only their real part.
\par
It is natural to assume that in the mixed quantum state the
readings of the detector such as
\begin{equation}\label{9}
{\rm Re}\langle u|A|v\rangle=\sum_jA_jw_j(u,v)\cos\Phi_j
\end{equation}
are averaged out to zero because of the random phase. In other words,
the measurements in the mixed quantum state give values of $A_j\cos\Phi_j$
(that may be positive or negative) with the probabilities
depending on the phase. For a random initial phase $\phi$ the phases
$\Phi_j=\alpha_j+\phi$ remain random and ${\rm Re}\langle
u|A|v\rangle$ vanishes after averaging over series of measurements:
$$\overline{\cos\Phi_j}=0.
$$ Note that in Eqs.~$(\ref{6})-(\ref{7})$ for the
$\overline{AB}$ correlation there is no phase difference and the
pairs from the different sources always give a contribution
irrespective of the phase $\Phi_j$. This cancelation of the phases
looks like a cancelation of the initial phase $\varphi$ in
Eqs.~(\ref{1}) or (\ref{2}).
\par
Thus the physical picture of the correlation of quantum particles
emitted by two independent sources and registered by two detectors
looks as follows. The sources produce uncorrelated quantum
particles (wavicles). These wavicles produce no correlation in the
measuring devices. Using the popular word "entanglement"\, one can
say that they are not entangled. The data obtained from such
particles remain the same either after separate or after joint
averaging. By separate averaging we mean the independent averaging
of the data of each particular detector.
\par
The source weavicle is a pair of {\em bra}\,+\,{\em ket} with
arbitrary phases of the opposite signs. As one can see above the
quantum correlation occurs when the uncorrelated wavicles exchange
their {\em bra} or {\em ket} and become new wavicles mutually
correlated due to the same phase difference of the opposite signs.
Just these correlated wavicles may be called the entangled quantum
particles. {They behave in a different way under separate and
joint measurements}. Their contributions vanish under the separate
averaging thus looking as a sort of noise for a detector. However,
due to their common phase these contributions {\em can be
redeemed by the joint averaging.} The common phase being the
real reason of data correlation does not enter into the final
expression for the quantity $\overline{AB}$. Thus the wave
function phase determines the result of measurements but is absent
in the final result. Therefore it may be called the actual hidden
variable of correlation.
\par
The random phase noise mentioned above can be measured and
analyzed. We believe that predictions concerning properties of
such a quantum noise can make a new feature as a part of the
theory of quantum entanglement.
\par
Now let us consider some correlation phenomena. For the intensity
correlation of two wave sources (known as the Hanbury
Brown-Twiss-effect [\cite{HBT}]) the operators $A$ and $B$ are the
space position operators at points ${\bf r}_1-{\bf r}_2={\bf R}$
and the wave functions are the plane waves $u\Rightarrow{\bf p}$,
$v\Rightarrow{\bf p'}$ ($\bf p$ and $\bf p'$ being the wave
vectors). Then the correlation is given by
\begin{equation}\label{10}
\overline{AB}=[1\pm \cos{\bf(p-p')R}]2F_{\bf p}F_{\bf p'}.
\end{equation}
Here the first term corresponds to uncorrelated weavicles creating
the homogeneous background while the second term is the
interference contribution of the wavicles created by the {\em bra}
and {\em ket} of different sources. Eq.~(\ref{10}) describes also
the bunching of bosons and anti-bunching of fermions while the flow
of them falls on a detecting screen. We see that for
$${\bf R\equiv (r_1-r_2)\rightarrow 0}$$ the fermions "avoid"\, each other
while the probability to find two bosons in one point becomes
twice as big as compared to the background value.
\par
For the Bohm version of EPR-paradox [\cite{EPR}] two spin measurements are
relating to two space points as well as to two time moments $t$
and $t'$. The observable spin values are time-independent with the
same result as for $t=t'$. Because of importance of this case
for the correlation interpretation we consider it in more details.
\par
The operators $A$ and $B$ now represent the spin values measured
at given directions. Such an operator is the scalar
product $S$ of the spin vector $\bf \sigma$ formed by the Pauli
matrices and the unit vector  $\bf n$ of a certain direction. In the
polar coordinates it is given by
$$S\equiv({\bf n,{\sigma}})=\sin\theta
\cos\varphi\sigma_x+\sin\theta\sin\varphi\sigma_y+\cos\theta\sigma_z.$$
The source wave functions $|u\rangle\equiv|\uparrow\rangle$ and
$|v\rangle\equiv|\downarrow\rangle$ now are the eigenfunctions of
$\sigma_z$ with the eigenvalues $s_u=1$ and $s_v=-1$.
\par
For pure spin states only $\sigma_z$ contributes to the matrix
elements of the operator $S$ and we have
\begin{eqnarray}\label{11}
\langle\uparrow|S|\uparrow\rangle=
\cos\theta\langle\uparrow|\sigma_z|\uparrow\rangle=\cos\theta,\\\nonumber
\langle\downarrow|S|\downarrow\rangle=
\cos\theta\langle\downarrow|\sigma_z|\downarrow\rangle=-\cos\theta.
\end{eqnarray}
For the mixed states only $\sigma_x$ and $\sigma_y$ give results so we have:\\
\begin{eqnarray}\label{12}
\langle\uparrow|S|\downarrow\rangle=
\cos\varphi\sin\theta\langle\uparrow|\sigma_x|\downarrow\rangle+
\sin\varphi\sin\theta\langle\uparrow|\sigma_y|\downarrow\rangle=
e^{i\varphi}\sin\theta,\\\nonumber
\langle\downarrow|S|\uparrow\rangle=
\cos\varphi\sin\theta\langle\downarrow|\sigma_x|\uparrow\rangle+
\sin\varphi\sin\theta\langle\downarrow|\sigma_y|\uparrow\rangle
=e^{-i\varphi}\sin\theta.
\end{eqnarray}
\par
Let the directions of spin measurements of the detectors $A$ and
$B$ be given by the unit vectors $\bf a$ and $\bf b$. Then using
Eqs.~(\ref{11}) and (\ref{12}) for the matrix elements and following
Eq.~(\ref{3}) we get:
\begin{eqnarray}\label{13}
\overline{A}=\langle\uparrow|A|\uparrow\rangle
F_\uparrow+\langle\downarrow|A|\downarrow\rangle F_\downarrow
=\cos\theta_a (F_\uparrow-F_\downarrow),\\\nonumber
\overline{B}=\langle\uparrow|B|\uparrow\rangle
F_\uparrow+\langle\downarrow|B|\downarrow\rangle F_\downarrow
=\cos\theta_b (F_\uparrow-F_\downarrow).\\\nonumber
\end{eqnarray}
Note that for $F_\uparrow=F_\downarrow$ the mean values of
$\overline{A}$ and $\overline{B}$ are zero irrespective of the
measurement directions. For the quantity $\overline{AB}$ as before
we have two contributions. The first one is given by
Eq.~(\ref{4})$$(\overline{AB})_{\rm uncor}=
-\cos\theta_a\cos\theta_b2F_\uparrow F_\downarrow.$$
We see that the product of two spin measurements of polarized
spins of both sources enters into the detectors in such a way that
each detector measures the spin of one of the sources. The second term
corresponds the expression (\ref{6}) where each detector measures
spin in the mixed state created by {\em bra}+{\em ket} of both
sources i.e. the spin of correlated wavicles. We have for this
correlated contribution
\begin{equation}\label{14}
(\overline{AB})_{\rm cor}=\mp\cos(\varphi_a-\varphi_b)\sin\theta_a\sin\theta_b2F_\uparrow F_\downarrow.
\end{equation}
Finally for the spin correlation we have
\begin{equation}\label{15}
\overline{AB}=-\cos\gamma 2F_uF_v
\end{equation}
where $\gamma$ is the angle between the measured spin directions
$\bf a$ and $\bf b$:
\begin{equation}\label{16}
\cos\gamma=\cos\theta_a\cos\theta_b+\cos(\varphi_a-\varphi_b)\sin\theta_a\sin\theta_b.
\end{equation}
We see that the sum of two contributions becomes independent of
the direction of the spin vector ${\bf \sigma}$ being proportional
to the scalar product of the unit vectors $\bf a$ and $\bf b$
which are the directions of independent spin observations in two
space points in one time moment.
\par
Now that we understand the physical cause of quantum correlation,
the dependence $(\ref{15})$ does not look puzzling and mysterious
even though the measurement directions $\bf a$ and $\bf b$ are
arbitrary and according to the quantum mechanics the spin values
do not exist before the measurements.
\par
For the case where $\overline{AB}$ is the matrix
element $\langle\Psi|AB|\Psi\rangle$ with the singlet wave
function $\Psi$, namely
\begin{equation}\label{17}
\Psi=|\uparrow_a\rangle|\downarrow_b\rangle-|\uparrow_b\rangle|\downarrow_a\rangle
\end{equation}
we have $$\overline{AB}=\langle\Psi|AB|\Psi\rangle,\quad
\overline{A}=\langle\Psi|A|\Psi\rangle=0,\quad
\overline{B}=\langle\Psi|B|\Psi\rangle=0$$ that corresponds to our result
of two fully independent sources under condition
\begin{equation}\label{18}
F_\uparrow=F_\downarrow=1.
\end{equation}
We see that the singlet wave function (\ref{17}) actually describes the
flow of oppositely polarized spins of equal intensities. This
equality is a unique property of the measurements described by
matrix elements with the singlet wave function $\Psi$.
\par
One should also remember that the correlation is of statistical
nature and to observe it a series of measurements is required. The
law $$\overline{S}=\langle\uparrow|S|\uparrow\rangle
F_\uparrow=\cos\theta F_\uparrow$$ for a flow of polarized spins in
the $z$ direction measured at the angle $\theta$ can be realized as random
series of positive and negative units that appear with the
probabilities depending of $\theta$:
\begin{equation}\label{19}
\cos\theta=(+1)\cos^2\theta/2+(-1)\sin^2\theta/2=\cos^2\theta/2-\sin^2\theta/2.
\end{equation}
As we see from Eq.~(\ref{13}) for two spin flows of the opposite
signs the mean observed value is proportional to the
difference of intensities of the flows:
\begin{equation}\label{20}
\overline{S}=\langle\uparrow|S|\uparrow\rangle
F_\uparrow+\langle\downarrow|A| \downarrow\rangle
F_\downarrow=\cos\theta(F_\uparrow-F_\downarrow).
\end{equation}
An observation of spins in the singlet states gives the zero value
for any measurement direction just because of the parity of
intensities $F_\uparrow=F_\downarrow$ and is not due to the
special properties of a singlet state.
For $\theta=\pi/2$ the detector always gives
mean zero irrespective of the flow intensity. In this case
there will be equal average number of positive and
negative units fixed by the detector. (Note that the expressions (\ref{19})and (\ref{20})
can be used for the probability interpretation of $\cos\Psi$ in Eq.~(\ref{9}))
\par
In conclusion we again emphasize that observed physical quantities
should be connected with pairs {\em bra}+{\em ket} together with the
explicit introduction of occupation numbers and their phases. We
believe that the physical picture of correlation phenomena
based on such an approach would not to look puzzling.


\newpage
\begin{references}
\bibitem{Cav} N. Brunner, D. Cavalcanti, S. Pironio, V. Scarani, S. Wehner, \rmp {\bf86}, 419 (2014).
\bibitem{NPG} M. Navascues and D. Peres-Garcia, \prl,{\bf 109}, 160405 (2012).
\bibitem{WSG} S. P. Walborn, A. Salles, R. M. Gomes et al., \prl, {\bf 106}, 130402 (2011).
\bibitem{M} T. Maudlin, Am. J. Phys, {\bf 78}, 121 (2010).
\bibitem{R} M. D. Reid, P. D. Drummond, W. P. Bowen et al., \rmp, {\bf 81}, 1727 (2009).
\bibitem{AFOV}L. Amico, R. Fazio, A. Osterloh, V. Vedral, \rmp, {\bf 80}, 518 (2008).
\bibitem{Zr}W. H. Zurek, \rmp, {\bf 75}, 715 (2003).
\bibitem{RBH} J. M. Raimond, M. Brune, S. Haroche, \rmp, {\bf 73}, 565 (2001).
\bibitem{Zel} A. Zeilinger, Rev. Mod. Phys., {\bf 71}, S288 (1999).
\bibitem{HBT} R. Hanbury Brown and R. Q. Twiss, Nature (London) {\bf
177}, 27 (1956); {\bf 178}, 1046 (1956).
\bibitem{EPR} A. Einstein, B. Podolsky, N. Rosen, Phys. Rev., {\bf 47}, 777 (1935).

\end{references}
\end{document}